\documentclass[seceq]{ptptex}

\usepackage{graphicx}

\title{
Nucleation in a sheared Ising model: effects of external field%
}

\author{
Rosalind J. \textsc{Allen}$^1$, Chantal \textsc{Valeriani}$^1$ and Sorin \textsc{T{\u{a}}nase-Nicola}$^2$%
}

\inst{
1. SUPA, School of Physics, The University of Edinburgh, 
Mayfield Road, Edinburgh EH9 3JZ, UK.\\
2. University of Michigan, Physics Dept.,  450 Church St., Ann Arbor, MI 48109-1040, USA
}

\abst{
Simulations using the Forward Flux Sampling method have shown a nonmonotonic dependence of the homogeneous nucleation rate on the shear rate for a sheared two dimensional Ising model [R. J. Allen {\em{et al}}, arXiv:cond-mat/0805.3029]. For quasi-equilibrium systems ({\em{i.e.}} in the absence of shear), Classical Nucleation Theory (CNT) predicts the dependence of the critical cluster size and the nucleation rate on the external magnetic field. We investigate the behaviour of the sheared Ising model as a function of the external field. At low external field strength, the same nonmonotonic behaviour holds  and the peak in the nucleation rate is remarkably insensitive to the field strength. This suggests that the same external field-dependence holds for the enhancement of nucleation by shear at low shear rates and the suppression of shear at high shear rates. At high field strength, the nucleation behaviour is qualitatively different.  We also analyse the size and shape of the largest cluster in the transition state configurations, as a function of the external field. In the sheared system, the transition state cluster becomes larger and more elongated as the field strength decreases. We compare our results for the sheared system to the predictions of the CNT for the quasi-equilibrium case, and find that the CNT cannot easily be used to describe nucleation in the system under shear.
}

\begin{document}

\maketitle


Nucleation in driven systems is a widespread and important phenomenon that remains poorly understood. For "quasi-equilibrium" systems, in the absence of external driving, Classical Nucleation Theory (CNT) predicts the dependence of the nucleation rate and the critical cluster size on the degree of supersaturation and the interfacial tension of the growing cluster, for moderate values of the supersaturation. No such simple theoretical description is available for nucleation in driven systems. In this paper, we consider nucleation in a system under shear. In experimental sheared systems, nucleation is likely to be affected by a wide range of factors including transport of particles to the growing cluster, changes in cluster shape, shear-induced ordering of the cluster, polydispersity and shear-induced cluster breakup. Intuition suggests that nucleation will be suppressed for high enough shear rates. At low shear rates, some simulations and experiments have observed suppression of nucleation \cite{Butler,Blaak1,Blaak2}, while others have suggested that nucleation may be enhanced by the shear \cite{Haw,Gray,cerda,Ackerson}. 

In a recent paper, we studied nucleation in a sheared two-dimensional Ising model, for a fixed value of the external magnetic field \cite{jcp_submit}. Although this is a highly simplified model system, it nevertheless shows a nonmonotonic dependence of the nucleation rate as a function of the shear rate. The nucleation rate increases, apparently linearly, with the shear rate for weak shear, achieves a peak for shear rate ${\dot{\gamma}}_{\rm{max}}$, and decreases nonlinearly as the shear rate is further increased. These results were obtained using the Forward Flux Sampling rare event simulation method \cite{FFS,FFS2,FFS3}. Similar results were obtained recently by other authors for a sheared Lennard-Jones system \cite{cerda}. For the Ising model, we used modified shear algorithms, combined with an analysis of the transition state configurations for the sheared nucleation process, to analyse the physical mechanisms underlying the nonmonotonic dependence of the nucleation rate on the shear rate. We found that nucleation is suppressed at high shear rates because of shear-induced cluster breakup. The enhancement of nucleation by shear at low shear rates is due to a combination of shear-enhanced "single spin flip" cluster growth (possibly due to the formation of kinks in the growing cluster), and shear-induced coalescence of surrounding spins and small clusters with the growing cluster. 

In this paper, we investigate nucleation in the sheared Ising system as a function of the applied external field. The external field is an important control parameter because, in a quasi-equilibrium system ({\em{i.e.}} in the absence of shear), it controls the extent to which the initial state is metastable. This is a crucial parameter in the Classical Nucleation Theory (CNT), on which much of our understanding of nucleation in quasi-equilibrium systems is based \cite{Farkas,Becker,Volmer,Zeldovitch}. 
The CNT postulates that nucleation can be described as a one-dimensional barrier crossing problem, in which the order parameter is the size of the growing cluster of particles of the thermodynamically stable phase, surrounded by particles of the metastable phase. For the Ising model studied here, these "particles" are in fact spins on a lattice. The assumptions of the CNT are only expected to hold for low supersaturation; at high supersaturation, nucleation is expected to involve coalescence between multiple growing clusters and the CNT is expected to fail, even for the quasi-equilibrium case. For low supersaturation, the CNT assumes that one cluster grows with fixed shape, and that its interfacial tension takes the bulk value, independently of the cluster size or shape. It further assumes that the nucleating phase has the properties of the bulk (e.g. the free energy per particle is as in the bulk), and that growth of the cluster is by attachment of single particles from the surroundings. The free energy barrier for nucleation is determined by the balance between the favourable bulk free energy of the growing cluster and its unfavourable interfacial free energy. If we assume, for a two-dimensional Ising lattice, that the growing cluster is circular, then
\begin{equation}\label{eq:cnt}
\Delta F = - 2 \pi R^2 h + 2 \pi R \sigma
\end{equation}
where $\Delta F$ is the free energy as a function of the cluster radius $R$, $\sigma$ is the interfacial tension of the cluster and $h$ is the external field (we note that the free energy change on flipping one spin is $2h$ \cite{Onsager}). The top of the free energy barrier $\Delta F^* = \Delta F(R^*)$ occurs for $d(\Delta F) / dR = 0$; so that 
\begin{equation}\label{eq:cnt1}
R^* = \frac{\sigma}{2h} \qquad \qquad \Delta F^* = \frac{\pi \sigma^2}{2h} \qquad \qquad n^* = \frac{\pi \sigma^2}{4h^2}
\end{equation}
where $n^* \equiv \pi (R^*)^2$ is the number of spins in the cluster, at the top of the free energy barrier.  If instead the cluster is assumed to be a square with sides of length $L$, Eqs. (\ref{eq:cnt}) and (\ref{eq:cnt1}) would instead give:
\begin{equation}\label{eq:cnt2}
\Delta F = - 2 L^2 h + 4 L \sigma
\end{equation}
\begin{equation}\label{eq:cnt3}
L^* = \frac{\sigma}{h} \qquad \qquad \Delta F^* = \frac{2 \sigma^2}{h} \qquad \qquad n^* = \frac{\sigma^2}{h^2}
\end{equation}
where we now use $n^* \equiv (L^*)^2$. Finally, if the cluster grows as a rectangle with sides of length $L$ and $(1+\nu) L$, Eqs. (\ref{eq:cnt}) and (\ref{eq:cnt1}) should be replaced by
\begin{equation}\label{eq:cnt4}
\Delta F = - 2(1+\nu) L^2 h + 2 L (2+\nu) \sigma
\end{equation}
and
\begin{equation}\label{eq:cnt5}
L^* = \frac{\sigma (2+\nu)}{ 2h (1+\nu)} \qquad \qquad \Delta F^* = \frac{(2+\nu)^2 \sigma^2}{2(1+\nu)h} \qquad \qquad n^* = \frac{\sigma^2 (2+\nu)^2}{ 4h^2 (1+\nu)}
\end{equation}
using $n^* \equiv (1+\nu) (L^*)^2$. Applying transition state theory to the free energy barrier crossing process, CNT provides an expression for the nucleation rate $I$:
\begin{equation}\label{eq:cntrate}
I = \kappa \exp{\left[- \frac{\Delta F^*}{k_BT}\right]}
\end{equation}
where the exponential term describes the Boltzmann probability of finding the system at the top of the free energy barrier ({\em{i.e.}} CNT assumes that the cluster is created by an equilibrium fluctuation of the metastable phase, and that the system remains in quasi-equilibrium as the cluster grows). The prefactor $\kappa$ describes the dynamics of the trajectories at the top of the barrier \cite{Daan_book}. This prefactor (which has a power law dependence on the size of the critical cluster) can be calculated by solving the Master Equation for the attachment and detachment of particles to the growing cluster \cite{Farkas,Becker,Volmer,Zeldovitch}. 

The two-dimensional Ising model is an excellent test case for theories of nucleation, because its interfacial tension between phases of up and down spins is known analytically \cite{Onsager}:
\begin{equation}\label{eq:ons}
\sigma = 2J - k_BT \log\left[\coth\left(\frac{J}{k_BT}\right)\right]
\end{equation}
where $J$ is the spin-spin coupling constant. The Hamiltonian $H$ is given by 
\begin{equation}\label{eq:ham}
H = - \frac{h}{k_BT} \sum_i s_i - \frac{J}{k_BT}\sum_i\sum_j^{'} s_i s_j
\end{equation}
where $s_i = \pm 1$ is the spin of lattice site $i$ and the double sum is over nearest neighbour pairs, avoiding double counting. The CNT has been extensively tested for the non-driven two-dimensional Ising model \cite{Neves,Shneidman2}. Although it is clear that the nucleation rate follows the general form (\ref{eq:cntrate}), the critical cluster size dependence of the prefactor $\kappa$ is still a matter of debate. Other simulations of nucleation in the Ising model in two and three dimensions have found evidence that the largest cluster size may not be the only important order parameter \cite{Pan,Peters}. 

In this paper, we study nucleation in a two-dimensional Ising model which is subjected to an applied shear. We mainly work in the regime of small external fields, where the CNT might be expected to be accurate in the absence of shear. However, in the presence of shear, there is no reason why the CNT predictions should apply. The CNT is an equilibrium theory, which assumes that a well-defined free energy barrier exists,  that the system is in quasi-equilibrium as it crosses this barrier, and that the probability of being at the top of the barrier is given by the Boltzmann factor. These assumptions cannot be made for a driven system. Moreover, the CNT further assumes that the shape of the growing cluster does not change during the nucleation process, and that the cluster grows by addition of single particles. These assumptions are also unlikely to hold in the case of nucleation under shear. However, the issue of how the nucleation rate and the critical cluster size and shape vary with the applied external field remains extremely relevant.  The applied external field is a central control parameter of the system, which we expect to have a strong effect on the nucleation rate and mechanism. By investigating how nucleation in the sheared Ising system is affected by the external field, we hope to gain insight into the essential physics of the sheared nucleation transition. We hope that such insight may assist in the construction of theories for nucleation under shear. 

\section{The Simulation Model}

\begin{figure}[h]
\begin{center}
{\rotatebox{0}{{\includegraphics[scale=0.4,clip=true]{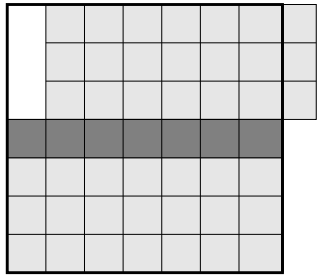}}}}
\caption{Illustration of a row shift in the shearing algorithm. The box is here of size $L=7$, and is shown surrounded by a thick black border. The row shaded grey has been chosen. All rows above the grey row have been shifted to the right. Periodic boundary conditions mean that the three lattice sites that appear to lie outside the simulation box are in fact in the left-most column. \label{fig:shear} }
\end{center}
\end{figure}

Our simulation model consists of a two-dimensional $L \times L$ square lattice of up-down spins, with Hamiltonian given by Eq.(\ref{eq:ham}). The coupling between nearest neighbour spins is $J=0.65 k_BT$. For this value of the coupling constant, the interfacial tension $\sigma \approx 0.74 k_BT$ \cite{Onsager}. Periodic boundary conditions are applied in the x and y directions.  We use external fields ranging from $h=0.045k_BT$ to $h=0.15 k_BT$.  Our simulations use Metropolis Monte Carlo (MC) dynamics. A single MC step consists of $L \times L$ attempted spin flips. In each attempted spin flip, a spin is chosen at random, the energy change $\Delta E$ on flipping is computed, and the spin flip is accepted with probability $P({\rm{acc}})=1$ if $\Delta E < 0$ and $P({\rm{acc}})=\exp{-\left[\Delta E / (k_BT)\right]}$ if $\Delta E > 0$. This choice of update rule does not preserve the total number of up spins, and does not model transport processes. The nucleation mechanism in Ising systems is sensitive to the update rule \cite{Binder_book}; it would be interesting to repeat this study for an update rule in which transport processes are accurately modelled, such as Kawasaki dynamics \cite{Kawasaki}.

We apply shear to the system using a modification of the algorithm proposed by Cirillo {\em{et al}} \cite{Cirillo,jcp_submit}. After every MC cycle, we make $N_s \times L$ attempts to shear the system. In each attempt, we choose a row at random and, with probability $P_s$, move all rows above this one to the right by one lattice site. The shear rate is given by ${\dot{\gamma}} = N_s P_s$. This algorithm is illustrated schematically in Figure \ref{fig:shear}, and is described in more detail in Ref. \cite{jcp_submit}. Care must by taken with the periodic boundary conditions in the y direction when a shear step is performed. As a result of a shear step, the "up"-neighbour of a spin in the top row is no longer the spin in the bottom row with the same x-index. In order to keep track of the identity of the "up"-neighbours of spins in the top row (and conversely the "down" neighbours of spins in the bottom row), we update a counter every time a shear step is performed. This counter can then be used to determine the necessary neighbour identities \cite{jcp_submit}.

\section{Forward Flux Sampling}

We compute nucleation rates and paths using Forward Flux Sampling (FFS) \cite{FFS,FFS2,FFS3}. This is a rare event simulation method which can be applied to equilibrium or nonequilibrium stochastic dynamical systems. Briefly, an order parameter $\lambda$ is defined which separates the initial and final states of the transition. If $\lambda < \lambda_A$, the system is in the initial (A) state, and if $\lambda > \lambda_B$, the system is in the final (B) state. The order parameter $\lambda$ is used to define a series of nonintersecting interfaces $\lambda_i$ for $1 < i < n-1$ in state space between the A and B states. The ``flux'' expression for the rate  $k_{AB}$ of transitions from A to B was first derived by Van Erp {\em{et al}} \cite{Titus}:
\begin{equation}
k_{AB} = \Phi_{A,0}\prod_{i=0}^{n-1} P( \lambda_{i+1} | \lambda_{i} )
\end{equation}\label{eq:ffs}
where we define $\lambda_0 \equiv \lambda_A$ and $\lambda_n \equiv \lambda_B$. Here, $\Phi_{A,0}$ is the flux of trajectories leaving the A state, and $P(\lambda_{i+1}|\lambda_{i})$ is the conditional probability that a trajectory which has reached interface $\lambda_i$ will subsequently reach the next interface $\lambda_{i+1}$ rather than returning to A. In FFS, $\Phi_{A,0}$ is computed with a simulation in the A state. Each time the interface $\lambda_0$ is crossed from A, the system configuration is stored. At the end of this run, the stored configurations are used to compute the probability $P(\lambda_1|\lambda_0)$ of reaching interface $\lambda_1$. A configuration from the collection is chosen at random and used as the starting point for a trial run which is continued until either $\lambda_1$ or $\lambda_0$ is reached. If the trial run reaches $\lambda_1$, its final configuration is stored in a new collection. This process is repeated a large number of times, to give an estimate of $P(\lambda_1|\lambda_0)$, and a new collection of configurations at $\lambda_1$. This collection is used to initiate trial runs to $\lambda_2$ (which are continued until either $\lambda_2$ of $\lambda_0$ is reached), and so on until the final interface $\lambda_n$ is reached. The rate constant is then calculated using Eq.(\ref{eq:ffs}). Transition paths (trajectories corresponding to transitions from A to B) can be extracted by tracing back successful trial paths from $\lambda_n$ back to $\lambda_0$.

\section{Nucleation rate as a function of shear rate}

\begin{figure}[h!]
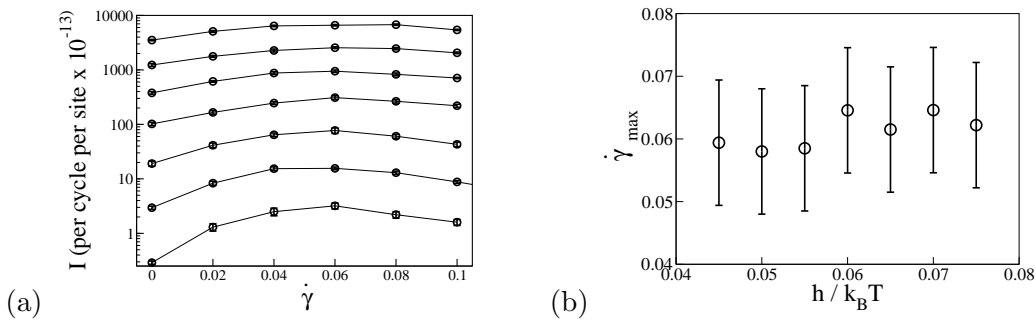

\begin{center}
\makebox[20pt][l]{(a)}{\rotatebox{0}{{\includegraphics[scale=0.22,clip=true]{65_all_rates.eps}}}}\hspace{1cm}\makebox[20pt][l]{(b)}{\rotatebox{0}{{\includegraphics[scale=0.22,clip=true]{maxima.eps}}}}
\caption{(a): Nucleation rate $I$ as a function of shear rate $\dot{\gamma}$, for several values of the external field. From bottom to top: $h=0.045 k_BT$, $h=0.05 k_BT$, $h=0.055 k_BT$, $h=0.06 k_BT$, $h=0.065 k_BT$, $h=0.07 k_BT$ and $h=0.075 k_BT$. (b): Positions of maxima of $I(\dot{\gamma})$ curves. \label{fig:rates}}
\end{center}
\end{figure}

In previous work \cite{jcp_submit}, we found a striking nonmonotonic dependence of the nucleation rate $I$ of the sheared Ising system on the shear rate $\dot{\gamma}$. For low shear rates, $I$ increases linearly with  $\dot{\gamma}$, but for shear rates $\dot{\gamma} > \dot{\gamma}_{\rm{max}}$, the nucleation rate decreases  nonlinearly. We showed that the suppression of nucleation at high shear rates is due to cluster breakup, while the enhancement of nucleation at low shear rates is due to both shear-enhanced ``single spin flip'' cluster growth, and coalescence of isolated spins and small clusters with the largest cluster. In this paper, we analyse how nucleation in the sheared Ising system is affected by the external field. Figure \ref{fig:rates}a shows  $I$ as a function of  $\dot{\gamma}$ for several different values of the external field $h$. Over the range $0.045 \le h \le 0.075$, the nucleation rate varies over four orders of magnitude, yet, remarkably, the nonmonotonic shape of the curves remains essentially unchanged. Figure \ref{fig:rates}b shows the position ${\dot{\gamma}}_{\rm{max}}$ of the peak in $I({\dot{\gamma}})$, plotted as a function of $h$. To within the statistical accuracy of our calculations, ${\dot{\gamma}}_{\rm{max}}$ does not vary with $h$. The position of this maximum is determined by the balance between enhancement of nucleation at low shear rates, and suppression of nucleation at high shear rates. Our observation that the peak position is independent of $h$  therefore indicates that the enhancement and suppression mechanisms must have the same functional dependence on the external field. Since we expect the size of the transition state cluster to depend on the external field strength, this leads us to the interesting hypothesis that the shear enhancement and suppression mechanisms have the same cluster size dependence.

\begin{figure}[h!]
\begin{center}
{\rotatebox{0}{{\includegraphics[scale=0.22,clip=true]{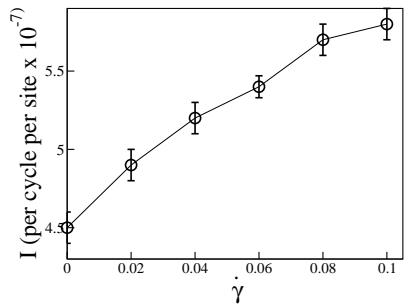}}}}\hspace{1cm}
\caption{Nucleation rate $I$ as a function of shear rate ${\dot{\gamma}}$, for a larger value of the external field, $h=0.15 k_BT$. \label{fig:supersat}}
\end{center}
\end{figure}

The results shown in Figure \ref{fig:rates} correspond to free energy barrier heights (in the absence of shear) of $15-24 k_BT$. In this range, we expect nucleation to involve one cluster of up spins. In the absence of shear, the nucleation mechanism changes significantly for larger values of $h$, where the free energy barrier is less than about $10k_BT$. Here, nucleation is expected to involve the simultaneous formation of many clusters of up spins (and the CNT is expected to fail, even in the absence of shear). We repeated our calculations for a larger value of the external field, $h=0.15 k_BT$, for which we expect a free energy barrier, in the unsheared system, $\Delta F^* \approx 7 k_BT$. Figure \ref{fig:supersat} shows that the nucleation rate as a function of shear is indeed very different at this external field value: the peak in the nucleation rate is apparently shifted to such a high shear rate that it is not reached in the range of ${\dot{\gamma}}$ simulated here.

\section{The Transition State Cluster}

To analyse the physical mechanism underlying nucleation in the presence of shear, we have extracted configurations from the Transition State Ensemble (TSE). The TSE is the ensemble of configurations belonging to the transition paths for which the value of the committor $P_B=0.5$ - {\em{i.e.}} trajectories fired from these configurations have equal probability or reaching the initial or final states. 
For external fields in the range $0.045 \le h \le 0.075$, as in Figure \ref{fig:rates}, we expect nucleation to proceed via the formation of a single cluster of up spins. We therefore analyse the size and shape of this largest cluster, for the TSE configurations, as a function of shear rate and external field strength. TSE configurations were obtained as described in Ref. \cite{jcp_submit}.

\begin{figure}[h!]
\begin{center}
{\rotatebox{0}{{\includegraphics[scale=0.22,clip=true]{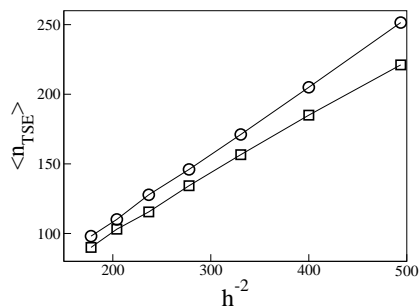}}}}\hspace{1cm}
\caption{Average size $\langle n_{\rm{TSE}} \rangle$ of largest cluster for TSE configurations, as a function of $h^{-2}$. Circles correspond to $\dot{\gamma}=0.0$; squares to $\dot{\gamma}=0.06$ \label{fig:size}}
\end{center}
\end{figure}

For quasi-equilibrium systems, the CNT predicts that the size $n^*$ of the critical cluster (the size of the largest cluster at the top of the free energy barrier), scales as $h^{-2}$, as in Eqs. (\ref{eq:cnt1}), (\ref{eq:cnt3}) and (\ref{eq:cnt5}). We therefore plot, in Figure \ref{fig:size}, the average number $\langle n_{\rm{TSE}} \rangle$ of up spins in the largest cluster for the TSE configurations, as a function of $h^{-2}$, for ${\dot{\gamma}}=0.0$ (no shear; circles) and ${\dot{\gamma}}=0.06$ (moderate shear; squares). For the zero shear case, we obtain the expected linear relationship, with slope $0.49 (k_BT)^2$, which lies in between the expected CNT results for a circular nucleus [$0.43 (k_BT)^2$; Eq(\ref{eq:cnt1})] and a square nucleus [$0.55 (k_BT)^2$; Eq(\ref{eq:cnt3})]. In the presence of shear, the relationship between $\langle n_{\rm{TSE}} \rangle$ and $h^{-2}$ appears also to be linear, but with a smaller slope of $0.41 (k_BT)^2$. We might imagine that the effect of shear is to elongate the growing nucleus. In this case, we might attempt to fit these results to a CNT-like picture with a rectangle-shaped nucleus, as in Eq.(\ref{eq:cnt5}). However, the resulting equation: $0.41 = \sigma^2 (2+\nu)^2/[4(1+\nu)]$, with $\sigma=0.74 k_BT$, has no real roots. We have also attempted to fit the nucleation rate $I$ to the CNT-like expression
\begin{equation}
\ln{I(h)} = a_0  + a_1 \ln{h} + \frac{a_2}{h}
\end{equation}
which arises from Eq.(\ref{eq:cntrate}), assuming a power-law dependence of the kinetic prefactor $\kappa$ on $h$. This results in a good fit, but with a parameter $a_2 \approx -0.4 \pm 0.1 $ that cannot be explained by CNT-like growth of a rectangular cluster as in Eq.(\ref{eq:cnt5}). Interestingly, this parameter does not appear to vary with shear rate for $\dot{\gamma} > 0.02$. These results suggest that although the nucleation rate and TSE cluster size have an apparently "CNT-like" dependence on the external field, this behaviour cannot easily be quantitatively explained by a CNT-like picture of the growing cluster.


\begin{figure}[h!]
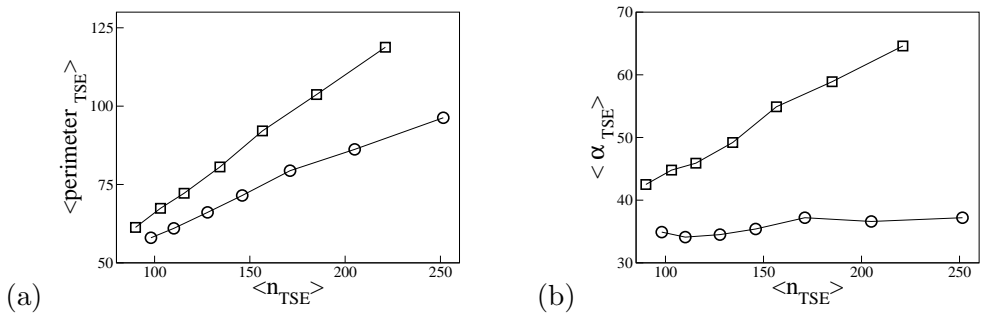

\begin{center}
\makebox[20pt][l]{(a)}{\rotatebox{0}{{\includegraphics[scale=0.22,clip=true]{perim_n.eps}}}}\hspace{1cm}\makebox[20pt][l]{(b)}{\rotatebox{0}{{\includegraphics[scale=0.22,clip=true]{shape_n.eps}}}}
\caption{(a): Average largest cluster perimeter  for the TSE configurations, plotted as a function of average cluster size. Each data point corresponds to a different external field strength.  (b):  Average value of $\alpha$, plotted as a function of average size, for TSE configurations at different external fields. Circles correspond to $\dot{\gamma}=0.0$; squares to $\dot{\gamma}=0.06$. \label{fig:shape}}
\end{center}
\end{figure}

We now turn to an analysis of  the shape of the largest cluster of up spins in the TSE configurations. Figure \ref{fig:shape}a shows the average perimeter of the largest cluster, defined as  the number of spins in the largest cluster with at least one down neighbour, as a function of the average cluster size $\langle n_{\rm{TSE}} \rangle$. Each data point corresponds to a different external field strength $0.045 \le h/(k_BT) \le 0.075$. In the absence of shear (circles), we expect the perimeter to scale as $n^{1/2}$, and indeed we see that the relationship is sub-linear. However, for moderate shear, $\dot{\gamma}=0.06$ (squares), the perimeter appears to scale linearly with the size of the largest cluster. This suggests that, in the presence of shear, larger TSE clusters (formed for small external field $h$) are more elongated  than smaller TSE clusters, formed at the same shear rate but at larger $h$. A dimensionless parameter describing the shape is $\alpha \equiv ({\rm{perimeter}})^2 / ({\rm{area}})$. Figure \ref{fig:shape}b shows the average value of $\alpha$ for the TSE configurations, as a function of the average cluster size. As expected, in the absence of shear, large and small transition state clusters (corresponding to small and large external fields) have the same shape. However, for moderate shear, larger transition state clusters (for small $h$), are more elongated than smaller ones (for large $h$). The relationship between $\langle \alpha \rangle_{{\rm{TSE}}}$ and $\langle n_{\rm{TSE}} \rangle$ is apparently linear, in agreement with the linear relationship between cluster perimeter and $\langle n_{\rm{TSE}} \rangle$ observed in Figure \ref{fig:shape}a.  These results imply that in the presence of shear, the shape of the TSE cluster is not independent of the applied external field. Smaller values of $h$ result in larger TSE clusters; these are more elongated than smaller TSE clusters which form for larger values of $h$. Although these results do not prove that a single cluster (at fixed $h$ and $\dot{\gamma}$) changes shape as it grows, they do suggest that this is possible. A generalisation of CNT for nucleation under shear would therefore need to include changes in cluster shape. 

\section{Conclusions}
In this paper, we have analysed nucleation in a sheared two-dimensional Ising system, as a function of the external field as well as of the shear rate. In previous work, we had found that the nucleation rate shows a peak at intermediate shear rate. Here, we find that this nonmonotonic behaviour is retained over a range of external field strength $h$ for which we expect nucleation to involve a single cluster of up spins. We also find that the position of the peak is remarkably insensitive to the external field. This suggests that the mechanisms behind the enhancement of nucleation at low shear rates and it suppression at high shear rates have the same $h$-dependence. We do not have a satisfactory explanation for this observation. When the external field is further increased so that the free energy barrier in the absence of shear is $\Delta F^* \approx 7 k_BT$, the behaviour changes qualitatively. It is known that for quasi-equilibrium systems, the single cluster picture breaks down for small free energy barriers. This appears also to be the case in the presence of shear.

We next analysed the size and shape of the transition state clusters, as a function of the external field strength (for moderate fields $0.045 \le h/(k_BT) \le 0.075$), in the absence of shear (${\dot{\gamma}}=0.0$) and for moderate shear (${\dot{\gamma}}=0.06$). The Classical Nucleation Theory (CNT) predicts a linear relationship between the critical cluster size and $h^{-2}$, and this was observed both in the presence and absence of shear. However, while the slope of this fit could be explained by CNT in the absence of shear, the slope for ${\dot{\gamma}}=0.06$ was not consistent with a CNT-like picture.  The shape of the transition state clusters was found to be elongated by the shear; in the presence of shear, larger TSE clusters (formed for small $h$) were more elongated than smaller TSE clusteres (formed for large $h$, at the same shear rate). The perimeter of the TSE clusters is linearly proportional to their area for ${\dot{\gamma}}=0.06$. Our results suggest that CNT cannot be applied to nucleation under shear without the relaxation of one of its core assumptions: fixed cluster shape.  This assumption could be relaxed by putting {\em{a priori}} knowledge of the cluster shape into the 
 "free energy" function, or perhaps by giving the "free energy"  a different functional dependence on the perimeter. Indeed the system does seem to behave as though its free energy depended more weakly on the perimeter than in CNT. However, it is important to bear in mind that for this nonequilibrium system the concept of free energy itself is anyway unclear.


In conclusion, this work shows that even for a highly simplified model system, nucleation under shear is a complex problem which requires the development of new theories. This work might take the form of attempting to reduce the problem to a one-dimensional dynamics in the largest cluster size, perhaps taking account of cluster shape, and including some mechanism for shear-mediated cluster coalescence and breakup (shown to be important in our previous work \cite{jcp_submit}). These theories should predict not only the effect of the external field, as investigated here, but also the role of the coupling constant, which we have not so far studied. Other interesting directions for future work would include analysing the nucleation behaviour of sheared Ising models where transport processes are modelled (for example, Kawasaki dynamics), and testing whether these findings apply to more complex model systems.

\section*{Acknowledgements}
The authors thank C. Dellago, D. Frenkel and P. R. ten Wolde for valuable discussions. Discussions were stimulated by a workshop at the Erwin Schroedinger Institute (Vienna).  R.J.A. was funded by the Royal Society of Edinburgh, and C.V. was funded by EPSRC under grant EP/E030173.

%

\end{document}